\documentclass[prc,twocolumn,aps,preprintnumbers,floatfix,nofootinbib]{revtex4}

\usepackage{graphicx, color, dcolumn, bm, amsmath}

\newcommand\diag{\operatorname{diag}}

\renewcommand\Re{\mathop{\mathrm{Re}}}

\newcommand\p{\partial}

\begin{document}

\title{
Femtoscopic Signature of Strong Radial Flow \\ 
in High-multiplicity $pp$ Collisions
}

\author{Yuji~Hirono}
 \email{yuji.hirono@stonybrook.edu}
\affiliation{
Department of Physics and Astronomy, Stony Brook University,
Stony Brook, New York 11794-3800, USA
}
\author{Edward~Shuryak}
\affiliation{
Department of Physics and Astronomy, Stony Brook University,
Stony Brook, New York 11794-3800, USA
}
\date{\today}

\begin{abstract}
Hydrodynamic simulations are used to calculate the identical pion HBT
radii, as a function of the pair momentum $k_{\rm T}$. 
This dependence is sensitive to the magnitude of the collective radial
flow in the transverse plane, and thus comparison to ALICE  data
enables us to derive its magnitude. By using hydro solutions with
variable initial parameters we conclude that in this case fireball
explosions start with a very small  initial size, well below 1 ${\rm fm}$.
\end{abstract}

\maketitle

\section{Introduction}

The so-called Hanbury-Brown-Twiss (HBT) interferometry method 
originally came from radio astronomy
\cite{brown1956correlation} as intensity interferometry. 
The influence of Bose symmetrization of the wave function of the
observed mesons in particle physics was first emphasized by Goldhaber et
al. \cite{Goldhaber:1960sf} and applied to proton-antiproton
annihilation. Its use for the determination of the size and duration of the
particle production processes had been proposed by Kopylov and
Podgoretsky \cite{Kopylov:1973qq} and one of us \cite{Shuryak:1974am}. 
Heavy-ion collisions, with their large multiplicities, turned  the
``femtoscopy'' technique into a large industry. Early applications for
RHIC heavy-ion collisions were in certain tension with the
hydrodynamical models, but this issue was later resolved; see, 
e.g., \cite{Pratt:2008qv}. The development of the HBT method had made it
possible to detect the magnitude and even deformations of the flow.

\begin{figure}[t]
\begin{center}
\includegraphics[width=60mm]{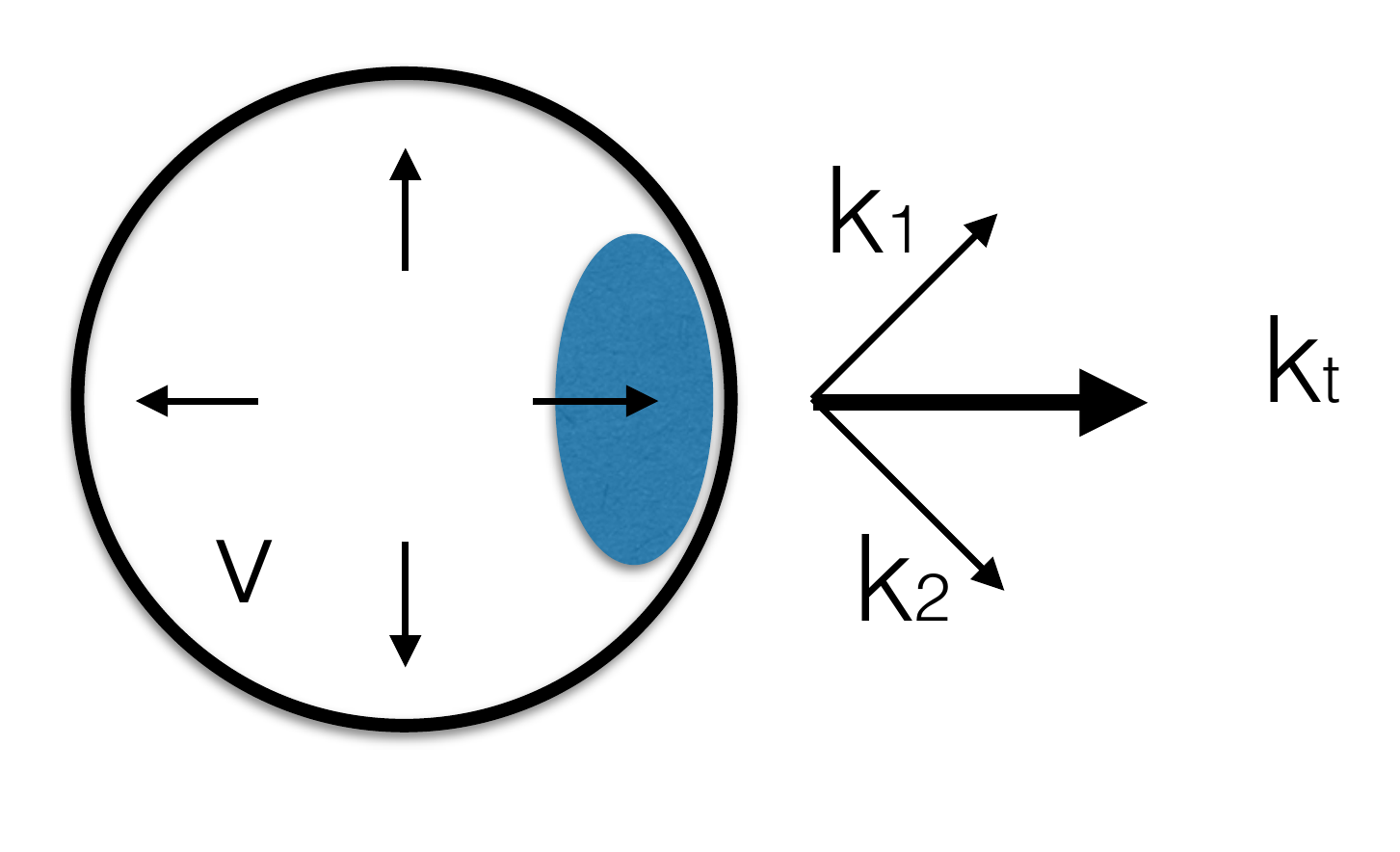}
\end{center}
\caption{ (Color online) Sketch of how the radial flow (arrows directed radially from the
 fireball center) influences the HBT radii. At small $k_{\rm T}$ the
 whole fireball (the circle) is visible, 
 but at larger $k_{\rm T}$ one sees only the
part comoving in the same direction, shown by the shaded ellipse.
}
\label{fig:sketch}
\end{figure}

Makhlin and Sinyukov \cite{Makhlin:1987gm} made the important
observation that HBT radii are sensitive to collective flows of matter.  
The radii decrease with the increase of the total transverse
momentum $\bm k_{\rm T} = (\bm p_{\rm 1T}+ \bm p_{\rm 2 T})/2$ of the pair.
A sketch shown in Fig.\ref{fig:sketch} provides a qualitative
explanation of this effect: 
 the larger is $k_{\rm T}$, the brighter becomes a small  (shaded) part
 of the fireball,  
the radial flow of which is maximal
 and its direction
 coincides with the direction of $\bm k_{\rm T}$. This follows from
 maximization of the Doppler-blue shifted thermal
 spectrum $\sim \exp\left( - p^\mu u_\mu/T_{\rm f}\right)$. 
 In this paper we will rely on this effect, as well as on ALICE HBT
 data, to deduce the magnitude of the flow in high multiplicity $pp$ collisions.

(Although we will not use those, let us also mention that the HBT method
can also be used not only for determination of the radial flow, but for
elliptic flow as well; see, e.g., early STAR measurements
\cite{Adams:2003ra}. Another development in the HBT field was
a shift from two-particle to three-particle correlations
\cite{Adams:2003vd},  \cite{Abelev:2014pja} available due to very high
multiplicity of events as well as high luminosities of RHIC and LHC colliders.) 

With the advent of the LHC it became possible to trigger on
high-multiplicity 
events, both in $pp$ and $pPb$ collisions: the resulting
sample revealed angular anisotropies $v_2,v_3$ similar to anisotropic
flows in heavy-ion ($AA$) collisions. At the moment the issue of whether
those can or cannot be  described hydrodynamically is under debate. 
  So far the discussion of the strength of the radial flow has been based 
on the spectra of identified particles; see
\cite{Shuryak:2013ke,Ghosh:2014eqa}.
In this paper we look at the radial flow from a different angle, 
using the measured HBT radii \cite{Aggarwal:2010aa}.

The HBT radii for $pp$ collisions at the LHC have been measured by the ALICE
Collaboration \cite{Aggarwal:2010aa}, as a function of multiplicity. 
Their magnitude has been compared to those coming from hydro modeling 
in Refs.~\cite{Shapoval:2013jca, Sinyukov:2012ut}. 
Our analysis of the HBT radii  focus on the strength of the radial flow. 
We illustrate how the radii, and especially the ratio $R_o /
R_s$, are indicative of the flow magnitude.

While at minimally biased collisions and small multiplicities the observed HBT radii are 
basically independent of the pair transverse momentum $k_{\rm T}$,  for high multiplicity
the observed radii decrease with $k_{\rm T}$. 
So, the  effect we are after appears only at the highest multiplicities -- the same ones which display
hydro-like angular correlations and modifications of the particle spectra. 
The
  strongest decrease, as expected, is seen 
 for  the so-called $R_{o}$ radius, for which this reduction  in the
 interval $k_{\rm T}= 0.1 - 0.7\, {\rm GeV} $ reaches about factor 4
in magnitude.

The $k_{\rm T}$ dependence of the HBT radii tells us about the strength
of the flow. 
The reason these data are quite important is the following: the HBT
radii at small $k_{\rm T}$  tell us the $final$ size of the fireball, at
the freezeout.  
The radii at large $k_{\rm T}$, 
combined with hydro calculations to be described below, can
shed light on the $initial$ size of the fireball, which we consider to
be the main result of this work. 

We do not speculate below on how such initial conditions can be created:
this should be determined by models of the initial state.  Our goal is
only to derive phenomenologically its parameters.  
Their importance stems from the fact that high-multiplicity $pp$
collisions create the most extreme conditions of matter density
reached so far.

\section{Method of analysis}

\subsection{Hydrodynamic evolution}

For heavy-ion collisions one has good command of the matter distribution
in nuclei, and thus can model the shape of the initial state rather accurately. 
However in the case of high-multiplicity $pp$ collisions -- which are 
certain fluctuations with small probability --
there is still no quantitative theory, and thus the shape 
remains unknown. 

A certain shape is preferable, not on physical but technical grounds.
 An analytic solution known as Gubser flow \cite{Gubser:2010ze}
 is restricted to a shape appearing in a stereographic projection from a sphere to the transverse plane.
Using the same shape had allowed us to compare our numerical solution to the corresponding analytic expression,
providing control of the code numerical accuracy.

In the Gubser solutions, the energy density and velocity take the form
\begin{equation}
\epsilon (\tau, r) = \frac{\epsilon_0 (2 q)^{8/3}}
{
\tau^{4/3} 
[
1+2q^2(\tau^2 + r^2) + q^4(\tau^2 - r^2)^2
]^{4/3}
},
\end{equation}
\begin{equation}
 v_{\perp}(\tau, r) = \frac{2 q^2 \tau r}{1 +q^2 \tau^2 +q^2 r^2}.
\end{equation}
The space-time characteristics of the system are parametrized by two variables,
\begin{equation}
\left(\, q \ [{\rm fm}^{-1}], \,\, \epsilon_0 \, \right). 
\end{equation}
(The parameter $q$ is widely used below, not to be confused with the
momentum transfer.) 
The dimensionless energy density parameter $\epsilon_0$ is related with the entropy
per unit rapidity as 
\begin{equation}
 \epsilon_0 = f_\ast^{-1/3} 
\left(
\frac{3}{16 \pi} 
\frac{dS}{d \eta}
\right)^{ 4/3}, 
\end{equation}
where $f_\ast = 11$ is the number of effective degrees of freedom in
QGP \cite{Gubser:2010ze}. 
The entropy per unit rapidity is inferred from the measured 
charged particle multiplicity, 
\begin{equation}
 \frac{dS}{d \eta} \simeq 7.5 \frac{d N_{\rm ch}}{d \eta}. 
\end{equation}
Thus, the values of $\epsilon_0$ can be fixed by charged particle multiplicity. 

\begin{figure}[htbp]
\begin{center}
\includegraphics[width=90mm]{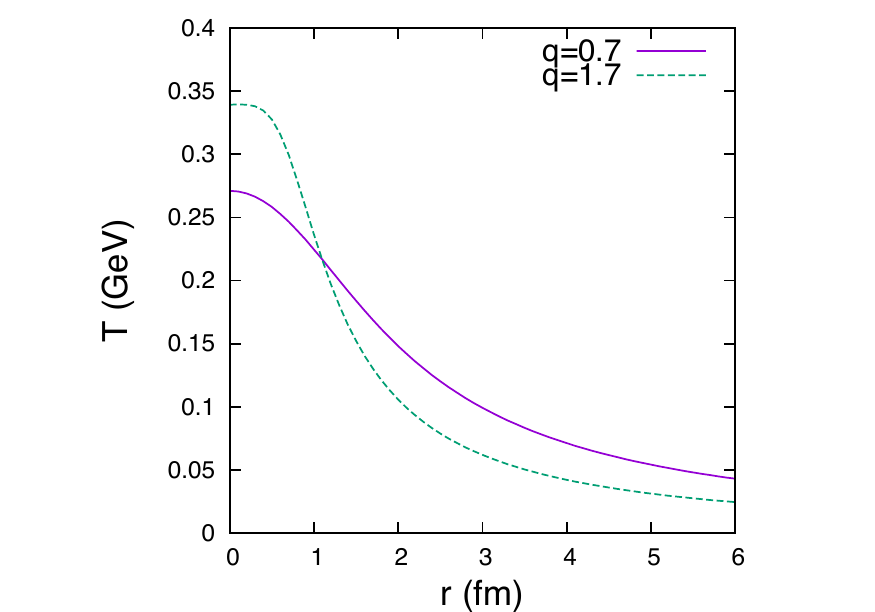}
\end{center}
\caption{ (Color online)
The temperature profiles of the Gubser solutions for different values
 of the parameter $q$, at $\tau=0.6 \ {\rm fm}$ as a function of $r$.
}
\label{fig:T-r}
\end{figure}

On the other hand, the parameter $q$ 
quantifies 
the size of the system. 
Figure \ref{fig:T-r} shows the temperature profiles at $\tau=0.6 \ {\rm fm}$ as a
function of $r$ for $q=1.7 \ {\rm fm}^{-1}$ and $q=0.7 \ {\rm fm}^{-1}$,
the ``smallest'' and ``largest'' fireballs in this study. 
One can see that the former fireball -- with larger $q$ -- is hotter and smaller in size. 

   While we use Gubser solution for early evolution of the system, 
unfortunately it  cannot be used  all the way to freezeout.
This solution was obtained by a conformal transformation and thus can only be 
used for conformal plasma with the conformal equation of state (EOS) $\epsilon=3p$. While it is believed to be a good approximation
for the early QGP phase of the collision, this is certainly not the case near the QCD phase transition,
where pressure $p$ remains roughly constant while 
the energy density $\epsilon$ changes by about an order of magnitude.  
Therefore, the initial Gubser-like stage is supplemented by a numerical
hydro solution, based on the realistic lattice-based EOS. 
We therefore start from the Gubser solution, but then,  at certain time $\tau_0 = 0.6 \ {\rm fm}$,  we
switch to numerical evolution with the realistic  EOS,  derived from recent lattice QCD
calculations \cite{Borsanyi:2013bia}. 

(We recall the ideal
relativistic hydrodynamic equations, 
\begin{equation}
 \p_\mu T^{\mu \nu} = 0,
\label{eq:eom}
\end{equation}
where $T^{\mu\nu}$ is the energy-momentum tensor. For a perfect fluid, 
$T^{\mu\nu}$ can be expressed as 
\begin{equation}
 T^{\mu\nu} = (\epsilon +p) u^\mu u^\nu - p \eta^{\mu\nu}, 
\end{equation}
where $\epsilon$ is the energy density, $p$ is the pressure, $u^\mu$ is
the fluid four-velocity, and $
\eta^{\mu \nu} \equiv \diag \{1,-1,-1,-1 \}
$ is the Minkowski metric. )

\begin{figure*}[tb]
\begin{center}
\includegraphics[width=150mm]{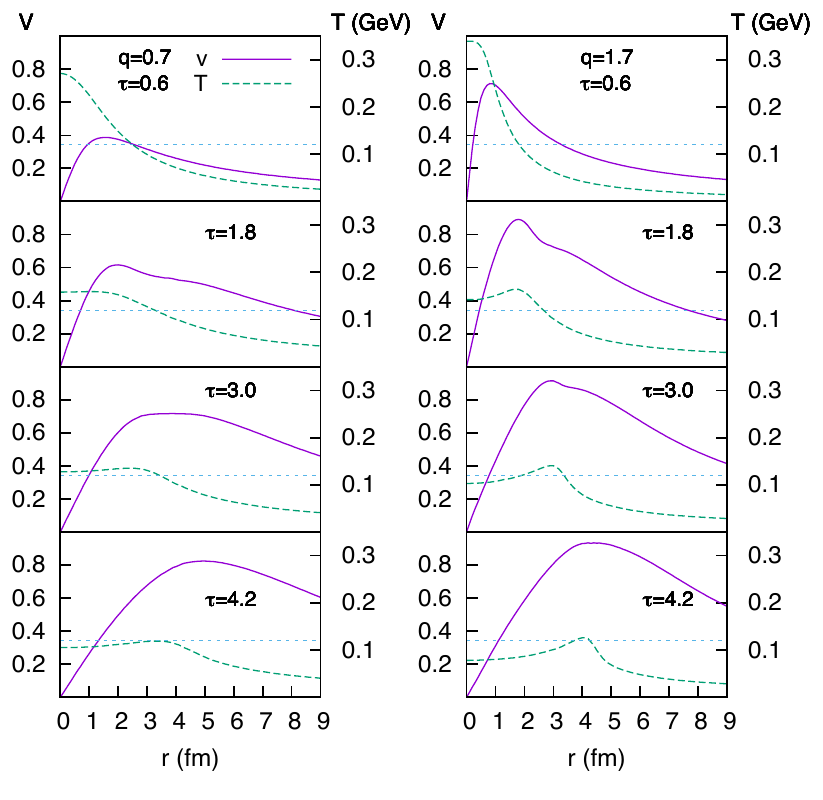}
\end{center}
\caption{ (Color online)
Time evolutions of temperature and velocity for $q=0.7 \ {\rm fm}^{-1}$ (left
 column) and $q=1.7 \ {\rm fm}^{-1}$ (right column). 
Temperature (dashed line) and velocity (solid line) profiles at $\tau =
 0.6, 1.8, 3.0, 4.2 \ {\rm fm}$ are plotted as a
 function of the radial coordinate $r$. 
The dotted lines in the plots indicate the freezeout temperature
 $T_{\rm f} = 0.12 \ {\rm GeV}$. 
}
\label{fig:evol}
\end{figure*}

\subsection{Freezeout}

In order to obtain the single-particle distribution from the hydrodynamic
solutions, we use the standard Cooper-Frye formula \cite{Cooper:1974mv}, 
\begin{equation}
p^0  \frac{d^3 N}{d \eta d p_{T}^2}
= \frac{1}{(2 \pi)^3}
\int \frac{ p^\mu d \sigma_\mu(x)}{\exp \left[ p \cdot u / T\right]
\mp_{\rm BF} 1}. 
\label{eq:cooper-frye}
\end{equation}
This formula is applied on a isothermal hypersurface characterized by
the freezeout temperature $T_{\rm f}$. 
We perform Monte-Carlo sampling of pions according to the distribution
(\ref{eq:cooper-frye}), following the steps below:
\begin{enumerate}
 \item Take a piece of surface elements $d \sigma^\mu$. We first calculated the average
       number of pions produced from this surface by 
\begin{equation}
 dN =  
\frac{1}{(2 \pi)^3}
\int 
\frac{d^3 p}{E}
\frac{ p^\mu d \sigma_\mu(x)}{\exp \left[ p \cdot u / T\right]
\mp_{\rm BF} 1}. 
\end{equation}
 \item 
Since $dN$ is typically a small number ($\sim 10^{-3}$), we can regard
       this number as a probability to produce a pion. 
According to this probability, we throw a dice and determine whether to make a pion or
       not.\footnote{
Although this treatment is justified for small $dN$, 
in general one should sample from the Poisson distribution with
mean $dN$. 
This method is applicable for larger surface elements from which more
       than one pion can be produced. 
}
If we are to produce a pion, we sample the momentum of the pion from the
       distribution
\begin{equation}
 f_1(x, \bm p) = 
\frac{1}{(2 \pi)^3}
\frac{ p^\mu d \sigma_\mu(x)}{\exp \left[ p \cdot u / T\right]
\mp_{\rm BF} 1}. 
\end{equation}
 \item We repeat the steps 1 and 2 for all the freezeout surface elements. 
\end{enumerate}
We refer the reader to Ref.~\cite{Hirano:2012kj} 
for the details of the sampling procedures.

\begin{figure*}[tb]
\begin{center}
\begin{minipage}[c]{0.32\hsize}
\includegraphics[width=80mm]{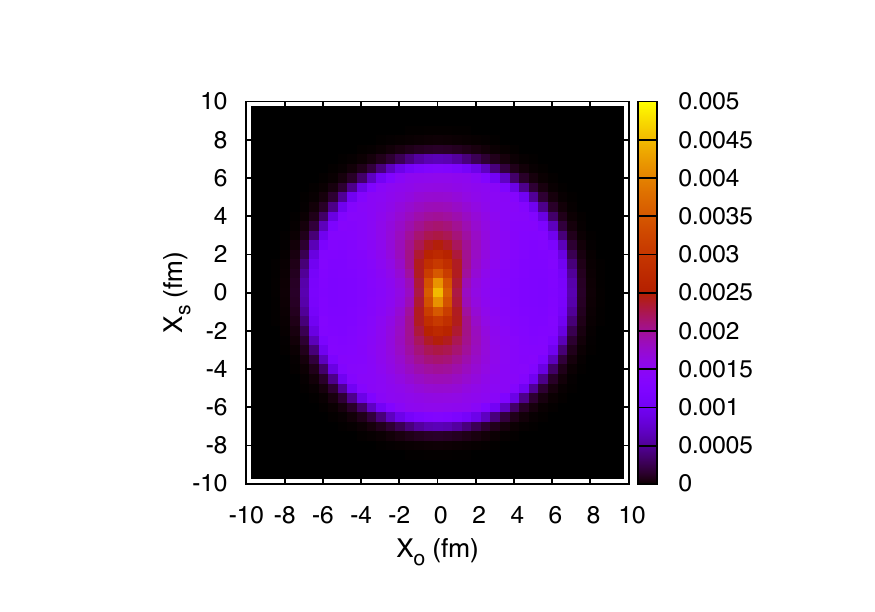}
{(a) $0.05 \ {\rm GeV} < k_{\rm T} < 0.15 \ {\rm GeV}$}
\end{minipage}
\begin{minipage}[c]{0.32\hsize}
\includegraphics[width=80mm]{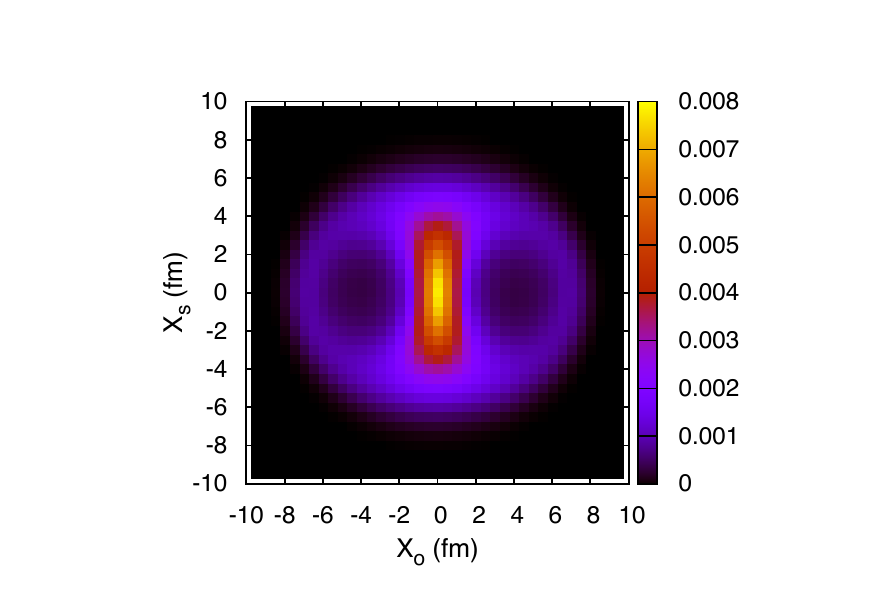}
{(b) $0.35 \ {\rm GeV} < k_{\rm T} < 0.45 \ {\rm GeV}$}
\end{minipage}
\begin{minipage}[c]{0.32\hsize}
\includegraphics[width=80mm]{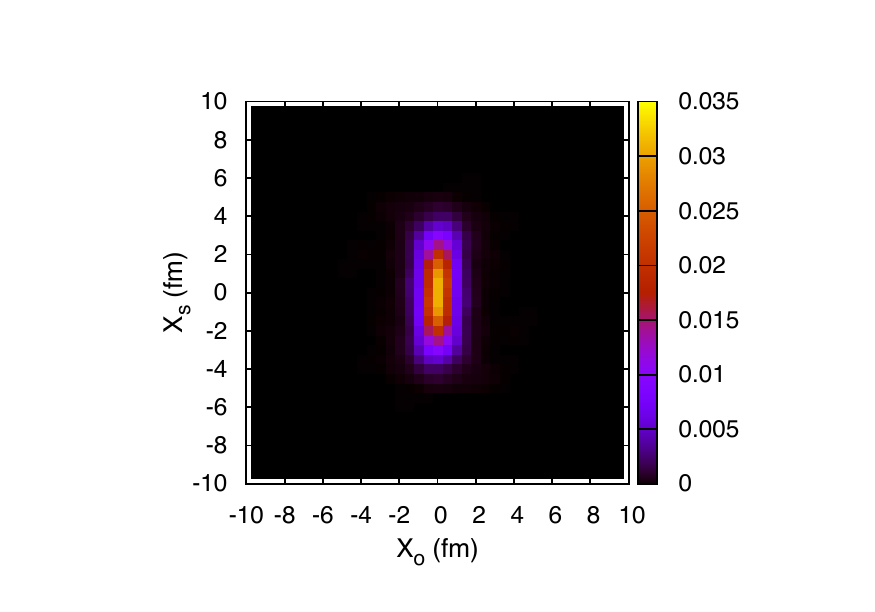}
{(c) $0.65 \ {\rm GeV} < k_{\rm T} < 0.75 \ {\rm GeV}$}
\end{minipage}
\caption{ (Color online)
Distribution of displacements in the out ($X_{\rm o}$) and side
 ($X_{\rm s}$) directions for $q=0.5 \ {\rm fm}^{-1}$. 
Three figures are for different $k_{\rm T}$ bins. 
}\label{fig:os_q=05}
\end{center}
\end{figure*}

\begin{figure*}[tb]
\begin{center}
\begin{minipage}[c]{0.32\hsize}
\includegraphics[width=80mm]{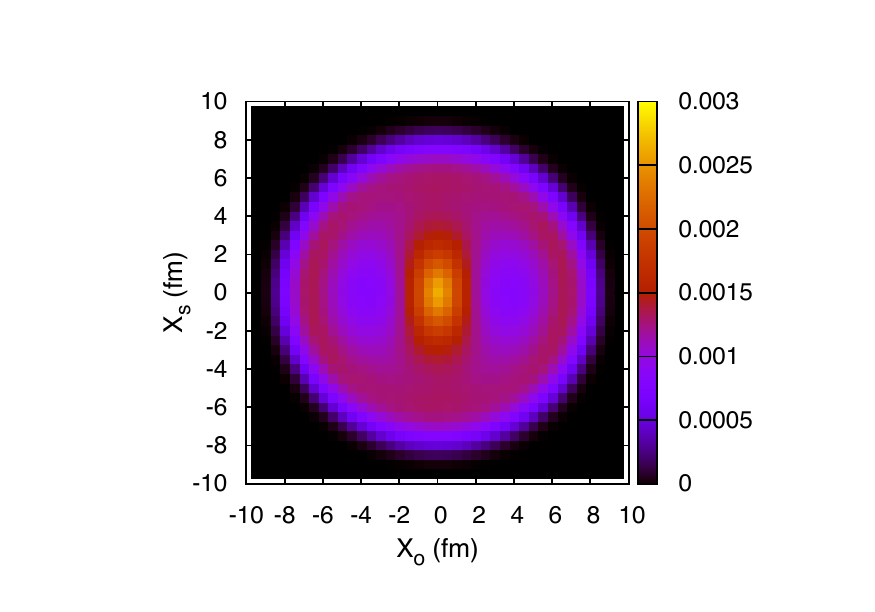}
{(a) $0.05 \ {\rm GeV} < k_{\rm T} < 0.15 \ {\rm GeV}$}
\end{minipage}
\begin{minipage}[c]{0.32\hsize}
\includegraphics[width=80mm]{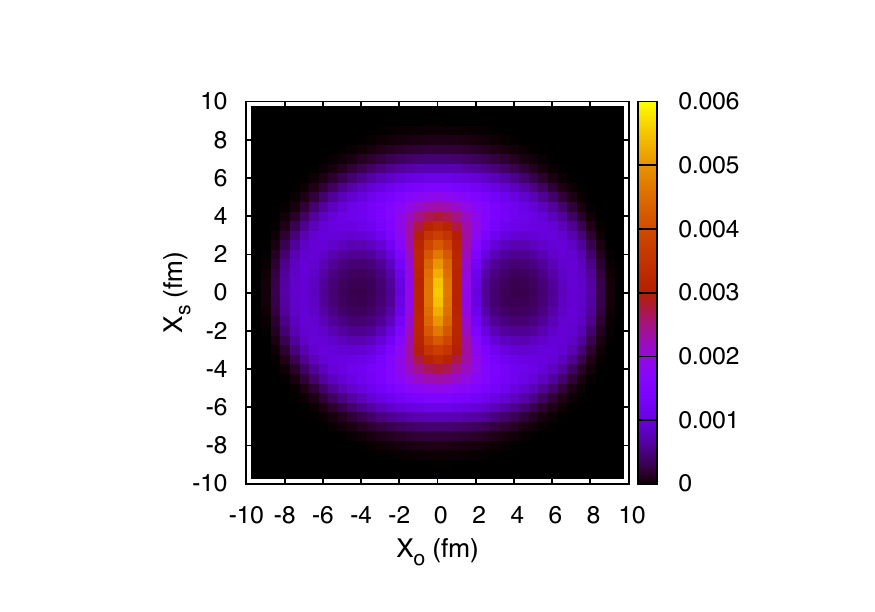}
{(b) $0.35 \ {\rm GeV} < k_{\rm T} < 0.45 \ {\rm GeV}$}
\end{minipage}
\begin{minipage}[c]{0.32\hsize}
\includegraphics[width=80mm]{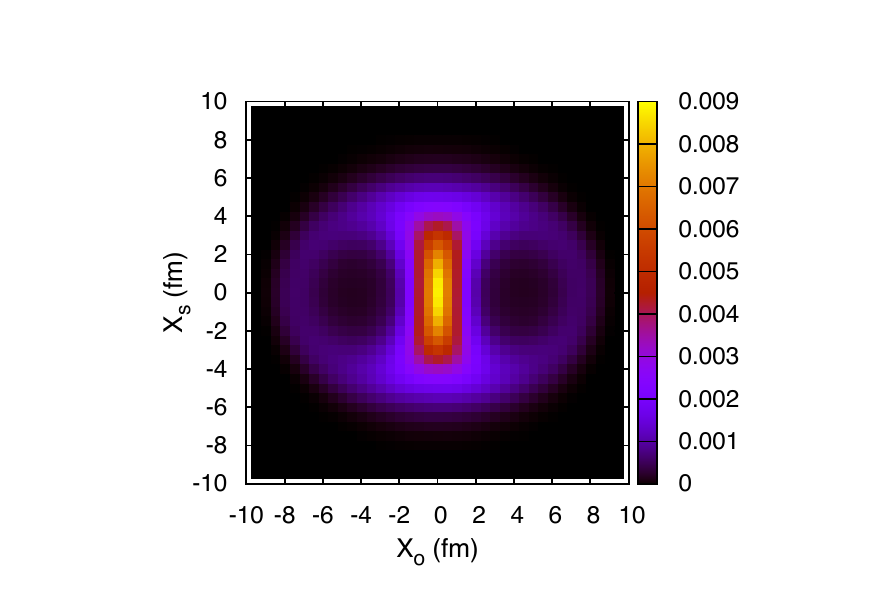}
{(c) $0.65 \ {\rm GeV} < k_{\rm T} < 0.75 \ {\rm GeV}$}
\end{minipage}
\caption{ (Color online)
Distribution of displacements in the out ($X_{\rm o}$) and side
 ($X_{\rm s}$) directions for $q=1.5 \ {\rm fm}^{-1}$. 
}
\label{fig:os_q=15}
\end{center}
\end{figure*}

\subsection{Calculations of correlations}

We have obtained the momenta and emission coordinates of produced
pions from the sampling based on the Cooper-Frye formula.
The effect of interference of identical particles is not
included at this stage, since the Cooper-Frye formula gives us only a {\it
single-particle} distribution function.  The {\it two-particle}
correlations come from Bose symmetrization
\begin{equation}
 C( k_{\rm T}, \bm q) 
= \frac{\sum_{< i, j> \in [k_{\rm T}]}\left[1 + \cos \left(q_\mu
 \Delta x^\mu\right) \right] 
}{ \sum_{< i, j> \in [k_{\rm T}]} 1 } , 
\label{eq:corr}
\end{equation}
where 
$\bm k_{\rm T}  \equiv (\bm p_{1{\rm T}} + \bm p_{2{\rm T}} )/2$ 
is the pair transverse momentum, 
$<i,j> \in [k_{\rm T}]$ indicates a pair of pions in a particular $k_{\rm T}$
bin, 
$q^\mu = p_1^\mu - p_2^\mu$ is four-momentum difference of a pion pair, 
and 
$\Delta x^\mu \equiv x_1^\mu - x_2^\mu$ is space-time distance of the pair. 
The correlation functions is evaluated in the ``longitudinally comoving frame'', 
where $k_z = 0$ for each pair. 
We impose a pseudo-rapidity cut $|\eta|<1.0$, by which the particles in the
mid-rapidity region are selected. 

We characterize the 3D correlation function in the 
``out-side-long'' parametrization \cite{Pratt:1986cc, Bertsch:1988db}, 
\begin{equation}
C(k_{\rm T}, \bm q) 
=1 + \lambda \exp\left[
- R_{o}^2 q_{o}^2 
- R_{s}^2 q_{s}^2 
- R_{\ell}^2 q_{\ell}^2 \right], 
\end{equation}
where 
$R_{o,s,\ell} =R_{o,s,\ell}( k_{\rm T})$ are the HBT radii of interest in this study, 
$q_{o}$ is the component of momentum parallel to the pair transverse momentum,
$q_{\ell}$ is the one parallel to the beam, 
and $q_{s}$ is the one perpendicular to out and long direction. 
For each $k_{\rm T}$ bin, we determined the
values of HBT radii by $\chi^2$ fitting.

\section{Results}

\subsection{Time evolution of fireballs}

The main qualitative feature of the solution is that the explosion 
 is stronger for smaller (hotter)  initial size -- or larger values of Gubser parameter $q$. 
Quantitatively 
 the time evolution of the temperature and
radial flow velocity for 
$q=0.7 \ {\rm fm}^{-1}$ (left column) and $q=1.7 \ {\rm fm}^{-1}$
(right column) is shown in
Fig.~\ref{fig:evol}.
The peak of the temperature in the central region $r\approx 0$ collapses, 
and the maximum moves to the rim of the fireball.
While the pressure gradient pushes out the matter,  the flow is increasing. 
One can see that the flow velocity reaches larger values for 
$q=1.7 \ {\rm fm}^{-1}$, compared to the case with $q=0.7 \ {\rm
fm}^{-1}$. 
Freezeout surfaces are located at the intersections of the
dashed lines (the fluid temperature)  and the dotted line (the assumed value 
of the freezeout temperature), where fluid
elements are turned into particles. 
At these intersections, the final flow is determined. 

We again emphasize that while the absolute freezeout times in both cases displayed is
similar ($\sim$ 4 fm), the flow
magnitude is quite different.  
As expected, it is significanly larger for smaller fireballs, or larger $q$.

\subsection{Flow and the distribution $P(\Delta x^\mu)$} 

Hydrodynamics gives us an intuitive explanation of the $k_{\rm T}$
dependence, as mentioned in the Introduction. 
If one selects a larger value of $k_{\rm T}$, the relevant region where
particles originate becomes smaller and more elliptic (see
Fig.~\ref{fig:sketch}). 
This intuitive picture can be quantitatively checked by looking at the
distribution, $P(\Delta x^\mu)$, of the pair-displacement vector
$\Delta x^\mu  = x_1^\mu - x_2^\mu$ and its $k_{\rm T}$ dependence. 

In Figs.~\ref{fig:os_q=05} and Fig.~\ref{fig:os_q=15}, 
we show the probability distribution of the displacement in ``out''
and ``side'' directions, $P(X_{o}, X_{s})$, for three 
$k_{\rm T}$ bins for two value of $q$ ($0.5 \ {\rm fm}^{-1} $ and $1.5 \ {\rm fm}^{-1}$). 
It is determined after the particle pairs are selected, from the Cooper-Frye
integral over the freezeout surface.
Here, $X_{o}$ is the projection of the displacement vector
$\Delta x^\mu$ to the direction of $\bm k_{\rm T}$, 
and $X_{s}$ is the projection of
$x^\mu$ in the direction perpendicular to $\bm k_{\rm T}$ and the beam axis. 
At low $k_{\rm T}$ [Fig.~\ref{fig:os_q=05}(a) and Fig.~\ref{fig:os_q=15}(a)], 
the distribution is broad and circular in out and side
directions. 

The wide circular component comes from the times when flow is still small, while
a narrow strip comes from the region where it is substantial. 
For higher $k_{\rm T}$, the distribution is squeezed, and is
narrower in the out direction compared to the side direction. 
These plots illustrate effect of the radial flow schematically shown in Fig.~\ref{fig:sketch}.

\subsection{HBT radii}

\begin{figure}[htbp]
\begin{center}
\includegraphics[width=80mm]{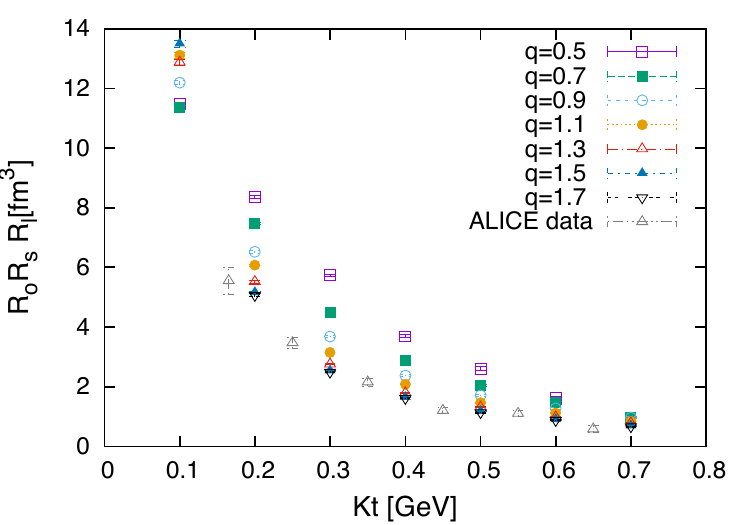}
\end{center}
\caption{ (Color online)
HBT volume as a function of the pair transverse momentum $k_{\rm T}$ for
 various values of the parameter $q$. 
$dN_{\rm ch}/d \eta = 27$, $T_{\rm f} = 120 \ {\rm MeV}$. 
}
\label{fig:v-kt}
\end{figure}

Now let us turn to the results of HBT radii. 
In Fig.~\ref{fig:v-kt}, we show the HBT ``volume`` ($R_{o}R_{s}R_{\ell}$) 
as a function of $k_{\rm T}$ for different values of $q$, together
with the experimental data from ALICE. The parameter $\epsilon_0$ is chosen to 
match the 
observed multiplicity in ALICE.
The radii from $q = 1.5$ to $1.7 \ {\rm fm}^{-1}$ reproduce the volume in the
ALICE data well. 

\begin{figure}[htbp]
\begin{center}
\includegraphics[width=80mm]{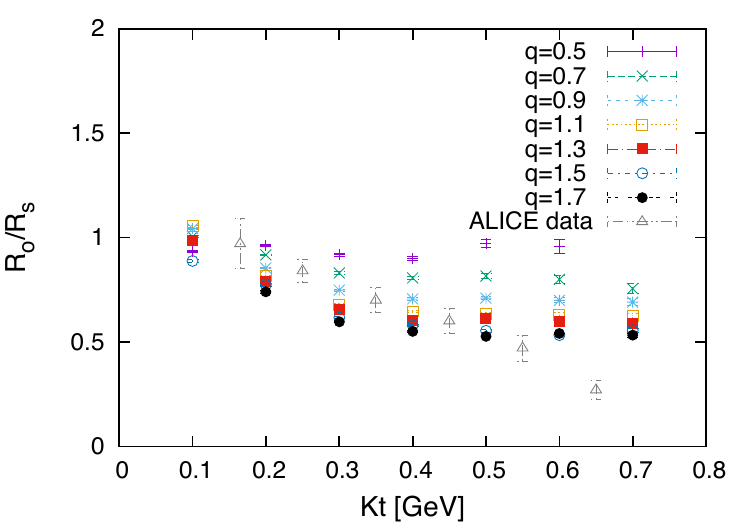}
\end{center}
\caption{ (Color online)
The ratio $R_{o}/R_{s}$ as a function of $k_{\rm T}$ for various
 values of $q$. The initial size is more compressed for a larger $q$,
 which results in stronger radial flow at freezeout. 
}
\label{fig:RoOverRs-kt}
\end{figure}

In Fig.~\ref{fig:RoOverRs-kt}, we show the ratio $R_o / R_s$ as a
function of $k_{\rm T}$. 
Basically, $R_o / R_s$ is a decreasing function of $k_{\rm T}$. 
At small values of $q$, the slope of $R_o / R_s$ is gentle. 
As $q$ becomes larger, the slope becomes steeper and 
$R_o / R_s$ is suppressed at large $k_{\rm T}$. 
The ALICE data shows further suppression compared to the result from the
largest value of $q$. 
Judging from the data, we can infer that $R_o / R_s$ is indicative of
the strength of the flow. 
However, the reason why $R_o / R_s$ is suppressed at large $k_{\rm T}$ is
not so trivial, which we explain in Sec.~\ref{sec:why}. 

\subsection{Why is the ratio $R_o/ R_s$ most sensitive to the strength of radial flow? }\label{sec:why}

Here we discuss the reason why $R_o / R_s$ is suppressed at large
$k_{\rm T}$ in the presence of a strong radial flow. 
Depending on the $k_{\rm T}$ cut, 
the area where particles originate changes. 
As $k_{\rm T}$ becomes higher, the region shrinks, especially in the
outward direction. 
If the system is composed of a gas with a large mean free path, such a
behavior would not be present. 
This trend indicates that the system is strongly interacting.
Furthermore, we claim that the ratio $R_o/ R_s$ is sensitive to 
the strength of the flow. 
What is difficult to understand is that, if one looks at the
distribution itself, $P(\Delta
x^\mu)$, 
the ratio of the widths of out and side directions, $L_o / L_s$, 
does not appear to be different for different values $q$ 
(compare Figs.~\ref{fig:os_q=05} and \ref{fig:os_q=15}). 

This might seem to be inconsistent with the behavior of $R_{o}/R_s$ at
large $k_{\rm T}$ calculated from the fitted radii: 
the ratio is almost unity at weak
flow case ($q=0.7 \ {\rm fm}^{-1}$), and it decreases as $q$ gets larger. 
Below we explain the reason for the apparent discrepancy. 
We will find that the suppression of the ratio $R_o/ R_s$ at large
$k_{\rm T}$ for the strong flow case is mainly driven by correlation of
emission time difference and distance of the emitted points in the out
direction. This 
was first pointed out in Ref.~\cite{Borysova:2005ng}
and  
is consistent with results in
Ref.~\cite{Kisiel:2010xy}, in which the HBT radii for $pp$ collisions
are studied using a blast-wave model.

We consider the quantities
\begin{equation}
\left. 
\frac{\p^2 C(k_{\rm T}, \bm q)}{\p q_i \p q_j} 
\right|_{\bm q = 0} , 
\end{equation}
where $i, j \in \{t, o,s,\ell \}$. 
When $P(\Delta x^\mu)$ is approximated by a Gaussian form, 
\begin{equation}
P(\Delta x^\mu) 
=
\frac{1}{ 16 \pi^2  V  }
\exp \left[
- \frac{X_t^2}{4 L_{t}^2} 
- \frac{X_o^2}{4 L_{o}^2} 
- \frac{X_s^2}{4 L_{s}^2} 
- \frac{X_\ell^2}{4 L_{\ell}^2} 
\right], 
\end{equation}
where $L_{t,o,s,l}$ are the widths in time, out, side, and long
directions, and $V \equiv L_t L_o L_s L_\ell$, 
the HBT radii can be
expressed by the moments as 
\begin{equation}
R_i^2
= 
\left. 
\frac{\p^2 C(k_{\rm T}, \bm q)}{\p q_i^2} 
\right|_{\bm q = 0}, 
\label{eq:radii-moments}
\end{equation}
with $i \in \{o,s,\ell \}$.

\begin{figure}[tbp]
\begin{center}
\includegraphics[width=80mm]{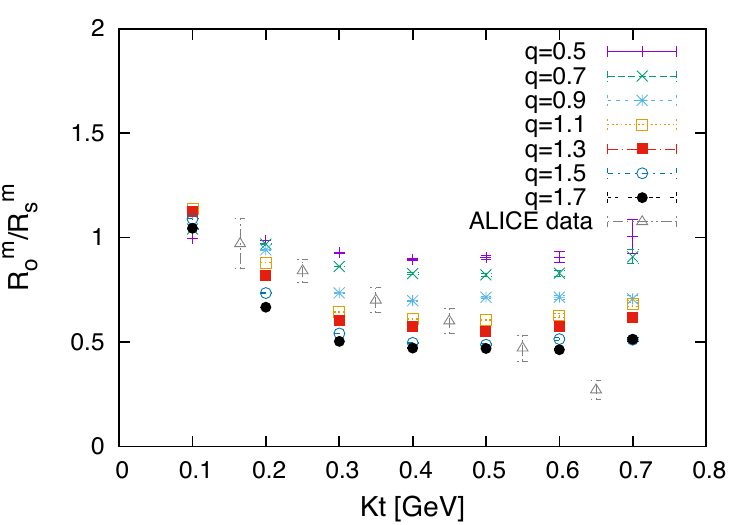}
\end{center}
\caption{ (Color online) $R_o / R_s$ calculated from the moments. 
}
\label{fig:ro_over_rs-kt}
\end{figure}

Below we express the measured HBT radii in terms of the
moments the distribution $P(\Delta x^\mu)$. 
The two-particle correlation function reads 
\begin{equation}
\begin{split}
C(k_{\rm T}, \bm q) - 1 
&= 
\int d X_t d X_o d X_s d X_\ell \ P( \Delta x^\mu) 
\cos \left(q_\mu \Delta x^\mu \right) \\
&= 
\Re \left[
\int d X_t d X_o d X_s d X_\ell \ P( \Delta x^\mu) 
e^{i q_\mu \Delta x^\mu}
\right]. 
\end{split}
\end{equation}
The exponent in the integral can be written as 
\begin{equation}
\begin{split}
 q_\mu \Delta x^\mu 
&= q_0 X_t - \bm q \cdot \Delta \bm x  \\
&= \bm \beta \cdot \bm q \  X_t  - \bm q_{\rm T} \cdot \Delta \bm x_{\rm T} \\
&= (\beta_{\rm T} q_{o} + \beta_{\rm L} q_\ell ) X_t  - q_{o} X_o - q_{s} X_s
 - q_\ell X_\ell, 
\end{split}
\label{eq:qx}
\end{equation}
where  $\bm \beta = \bm k / k_0$ and we used 
$\beta_s = 0$ ($\bm \beta $ is parallel to $\bm k$), and $q_0 = \bm
\beta \cdot \bm q$ ($ \leftrightarrow k_\mu q^\mu=0$), and $\beta_{\rm
T}$ and $\beta_{\rm L}$ are the projections of $\bm \beta$ in transverse
and longitudinal directions. 
In the current case, where the correlations function is evaluated in the
frame with $k_z=0$, $\beta_{\rm T} = \beta$ and $\beta_{\rm L}=0$. 
Thus, the HBT radii and the moments are related by 
\begin{eqnarray}
R_o^2 &= &
\left< 
\left(
X_o - \beta X_t
\right)^2
\right>  \nonumber \\
&=& 
\left< 
X_o^2
\right> 
+
\left< 
\beta^2 X_t^2
\right> 
- 2  
\left< 
\beta X_t X_o 
\right> , \label{eq:ro-moments}
 \\
R_s^2 &= &
\left< X_s^2 \right> , \label{eq:ro-moments-2}\\
R_\ell^2 &= &
\left< X_\ell^2 \right> . \label{eq:ro-moments-3}
\end{eqnarray}
Indeed, one can see that the radii calculated from the moments,  
using Eqs.~(\ref{eq:ro-moments}), (\ref{eq:ro-moments-2}) and 
(\ref{eq:ro-moments-3}), shows consistent behavior with 
the ones obtained by fitting procedure, compare 
Figs.~\ref{fig:ro_over_rs-kt} and \ref{fig:RoOverRs-kt}.

\begin{figure}[tbp]
\begin{center}
\includegraphics[width=80mm]{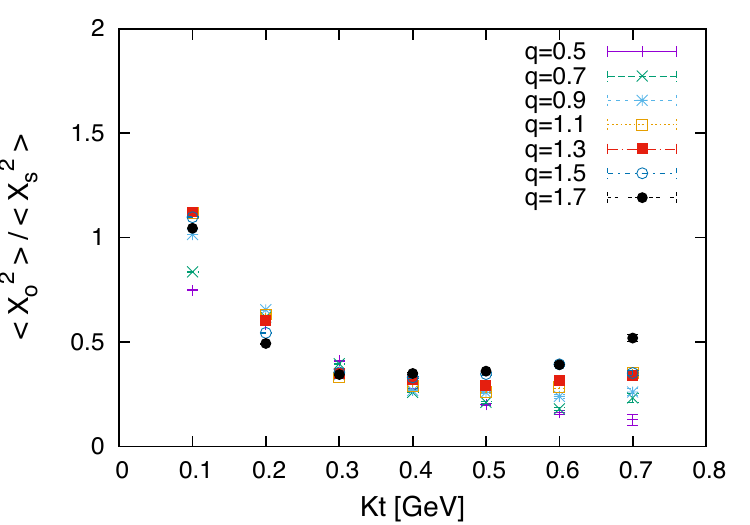}
\end{center}
\caption{ (Color online) 
$
\left<
X_o^2
\right> / 
\left<
X_s^2
\right> 
$ as a function of $k_{\rm T}$. 
}
\label{fig:xsq_out_over_rs2-kt}
\end{figure}

\begin{figure}[tbp]
\begin{center}
\includegraphics[width=80mm]{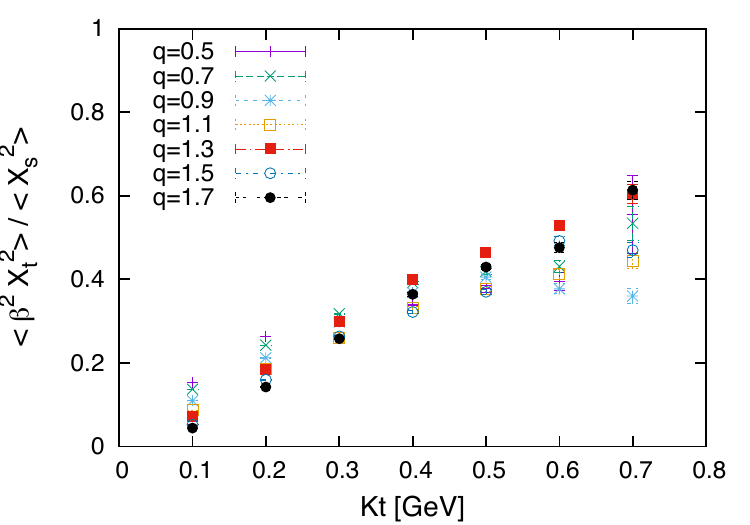}
\end{center}
\caption{ (Color online) 
$
\left<
\beta^2 X_t^2
\right> / 
\left<
X_s^2
\right> 
$
as a function of $k_{\rm T}$. 
}
\label{fig:betasq_xtsq_over_rs2-kt}
\end{figure}

\begin{figure}[tbp]
\begin{center}
\includegraphics[width=80mm]{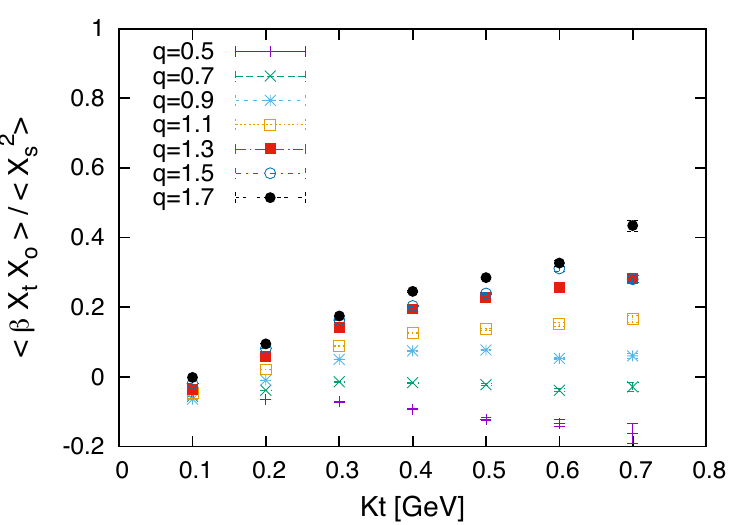}
\end{center}
\caption{ (Color online) 
$
\left<
\beta X_t X_o
\right> / 
\left<
X_s^2
\right> 
$
as a function of $k_{\rm T}$. 
}
\label{fig:beta_xt_xo_over_rs2-kt}
\end{figure}

Now let us discuss the reason why $R_o / R_s$ is suppressed at large 
$k_{\rm T}$ in the presence of strong flow. 
In terms of the ratio of moments, $R_o / R_s$ is composed of three terms,
\begin{equation}
\frac{R_o^2 }{R_s^2}
= 
\frac{ \left< X_o^2 \right> }{\left< X_s^2 \right>}
+
\frac{\left< \beta^2 X_t^2 \right> }
{\left< X_s^2 \right>}
- 2  
\frac{\left< \beta X_t X_o \right>}
{\left< X_s^2 \right>}
, 
\end{equation}
In order to see which term plays the dominant role in the suppression of
$R_s / R_o$ for strong flow, 
we plotted the values of each term for different values of the Gubser
parameter $q$, as a function of $k_{\rm T}$ 
(Figs.~\ref{fig:xsq_out_over_rs2-kt}, 
\ref{fig:betasq_xtsq_over_rs2-kt}, and 
\ref{fig:beta_xt_xo_over_rs2-kt}
). 
The behavior of the first term $ \left< X_o^2 \right> / \left< X_s^2
\right>$ is shown in Fig.~\ref{fig:xsq_out_over_rs2-kt}. 
For all the values of $q$, the ratio is around $1$ at lowest $k_{\rm
T}$, and is less than unity at higher $k_{\rm T}$. 
Note the fact that, at highest $k_{\rm T}$, the ratio is more suppressed
for weaker flows.
This indicates that the suppression of $R_o / R_s$ at large $k_{\rm T}$ 
for a strong flow is not caused by the term 
$ \left< X_o^2 \right> / \left< X_s^2 \right>$.

The suppression of $R_o / R_s$ is driven by 
$\left< \beta X_t X_o \right> / \left< X_s^2 \right>$, which is shown 
in Fig.~\ref{fig:beta_xt_xo_over_rs2-kt}. 
This term is a measure of correlation between emission time difference and
the displacement in the out direction. 
For a weak flow (small $q$), it is close to zero and the correlation is
weak for the entire region of $k_{\rm T}$. 
As we go to stronger flow (larger $q$), the lines rises and 
the correlation at high $k_{\rm T}$ becomes stronger. 
Since this term contributes to $R_o / R_s$ with a negative sign, 
it leads to the suppression of $R_o / R_s$ at large $k_{\rm T}$.


\section{Summary}

ALICE HBT data  \cite{Aggarwal:2010aa} provided a striking indication that
the highest multiplicity bin of $pp$ collisions at the LHC is rather
different from others: it shows evidence of strong radial flow. We
performed simulations of the system, using ideal relativistic
hydrodynamics. The early evolution is described by a Gubser conformal
solution, complemented by a numerical one, with a realistic EOS at later
stages. We show how strength of the radial flow depends on the initial
size and temperature of the fireball.

Comparison of the resulting HBT radii with high multiplicity data shows the best agreement
only for the smallest fireball we study, with Gubser parameter $q=1.5 - 1.7 \
{\rm fm}^{-1}$.   It confirms that
one in fact observes the presence of collective hydrodynamical flow in an 
unprecedented small system, smaller than 1 fm initially.  

\acknowledgements
Y.H. is grateful to T.~Kawanai, K.~Murase, and Y.~Tachibana for helpful discussions
regarding numerical implementations. 
Y.H. is supported by JSPS Research Fellowships for Young Scientists. 
The work of E.S. is supported in part by the U.S. Department of Energy
under Contract No. DE-FG02-88ER40388.

\bibliography{refs}

\begin{thebibliography}{22}
\expandafter\ifx\csname natexlab\endcsname\relax\def\natexlab#1{#1}\fi
\expandafter\ifx\csname bibnamefont\endcsname\relax
  \def\bibnamefont#1{#1}\fi
\expandafter\ifx\csname bibfnamefont\endcsname\relax
  \def\bibfnamefont#1{#1}\fi
\expandafter\ifx\csname citenamefont\endcsname\relax
  \def\citenamefont#1{#1}\fi
\expandafter\ifx\csname url\endcsname\relax
  \def\url#1{\texttt{#1}}\fi
\expandafter\ifx\csname urlprefix\endcsname\relax\def\urlprefix{URL }\fi
\providecommand{\bibinfo}[2]{#2}
\providecommand{\eprint}[2][]{\url{#2}}

\bibitem[{\citenamefont{Brown and Twiss}(1956)}]{brown1956correlation}
\bibinfo{author}{\bibfnamefont{R.~H.} \bibnamefont{Brown}} \bibnamefont{and}
  \bibinfo{author}{\bibfnamefont{R.}~\bibnamefont{Twiss}},
  \bibinfo{journal}{Nature} \textbf{\bibinfo{volume}{177}}, \bibinfo{pages}{27}
  (\bibinfo{year}{1956}).

\bibitem[{\citenamefont{Goldhaber et~al.}(1960)\citenamefont{Goldhaber,
  Goldhaber, Lee, and Pais}}]{Goldhaber:1960sf}
\bibinfo{author}{\bibfnamefont{G.}~\bibnamefont{Goldhaber}},
  \bibinfo{author}{\bibfnamefont{S.}~\bibnamefont{Goldhaber}},
  \bibinfo{author}{\bibfnamefont{W.-Y.} \bibnamefont{Lee}}, \bibnamefont{and}
  \bibinfo{author}{\bibfnamefont{A.}~\bibnamefont{Pais}},
  \bibinfo{journal}{Phys.Rev.} \textbf{\bibinfo{volume}{120}},
  \bibinfo{pages}{300} (\bibinfo{year}{1960}).

\bibitem[{\citenamefont{Kopylov and Podgoretsky}(1974)}]{Kopylov:1973qq}
\bibinfo{author}{\bibfnamefont{G.}~\bibnamefont{Kopylov}} \bibnamefont{and}
  \bibinfo{author}{\bibfnamefont{M.}~\bibnamefont{Podgoretsky}},
  \bibinfo{journal}{Sov.J.Nucl.Phys.} \textbf{\bibinfo{volume}{18}},
  \bibinfo{pages}{336} (\bibinfo{year}{1974}).

\bibitem[{\citenamefont{Shuryak}(1973)}]{Shuryak:1974am}
\bibinfo{author}{\bibfnamefont{E.~V.} \bibnamefont{Shuryak}},
  \bibinfo{journal}{Yad.Fiz.} \textbf{\bibinfo{volume}{18}},
  \bibinfo{pages}{1302} (\bibinfo{year}{1973}).

\bibitem[{\citenamefont{Pratt}(2009)}]{Pratt:2008qv}
\bibinfo{author}{\bibfnamefont{S.}~\bibnamefont{Pratt}},
  \bibinfo{journal}{Phys.Rev.Lett.} \textbf{\bibinfo{volume}{102}},
  \bibinfo{pages}{232301} (\bibinfo{year}{2009}), \eprint{0811.3363}.

\bibitem[{\citenamefont{Makhlin and Sinyukov}(1988)}]{Makhlin:1987gm}
\bibinfo{author}{\bibfnamefont{A.}~\bibnamefont{Makhlin}} \bibnamefont{and}
  \bibinfo{author}{\bibfnamefont{Y.}~\bibnamefont{Sinyukov}},
  \bibinfo{journal}{Z.Phys.} \textbf{\bibinfo{volume}{C39}},
  \bibinfo{pages}{69} (\bibinfo{year}{1988}).

\bibitem[{\citenamefont{Adams et~al.}(2004)}]{Adams:2003ra}
\bibinfo{author}{\bibfnamefont{J.}~\bibnamefont{Adams}} \bibnamefont{et~al.}
  (\bibinfo{collaboration}{STAR Collaboration}),
  \bibinfo{journal}{Phys.Rev.Lett.} \textbf{\bibinfo{volume}{93}},
  \bibinfo{pages}{012301} (\bibinfo{year}{2004}), \eprint{nucl-ex/0312009}.

\bibitem[{\citenamefont{Adams et~al.}(2003)}]{Adams:2003vd}
\bibinfo{author}{\bibfnamefont{J.}~\bibnamefont{Adams}} \bibnamefont{et~al.}
  (\bibinfo{collaboration}{STAR Collaboration}),
  \bibinfo{journal}{Phys.Rev.Lett.} \textbf{\bibinfo{volume}{91}},
  \bibinfo{pages}{262301} (\bibinfo{year}{2003}), \eprint{nucl-ex/0306028}.

\bibitem[{\citenamefont{Abelev et~al.}(2014)}]{Abelev:2014pja}
\bibinfo{author}{\bibfnamefont{B.~B.} \bibnamefont{Abelev}}
  \bibnamefont{et~al.} (\bibinfo{collaboration}{ALICE Collaboration})
  (\bibinfo{year}{2014}), \eprint{1404.1194}.

\bibitem[{\citenamefont{Shuryak and Zahed}(2013)}]{Shuryak:2013ke}
\bibinfo{author}{\bibfnamefont{E.}~\bibnamefont{Shuryak}} \bibnamefont{and}
  \bibinfo{author}{\bibfnamefont{I.}~\bibnamefont{Zahed}},
  \bibinfo{journal}{Phys.Rev.} \textbf{\bibinfo{volume}{C88}},
  \bibinfo{pages}{044915} (\bibinfo{year}{2013}), \eprint{1301.4470}.

\bibitem[{\citenamefont{Ghosh et~al.}(2014)\citenamefont{Ghosh, Muhuri, Nayak,
  and Varma}}]{Ghosh:2014eqa}
\bibinfo{author}{\bibfnamefont{P.}~\bibnamefont{Ghosh}},
  \bibinfo{author}{\bibfnamefont{S.}~\bibnamefont{Muhuri}},
  \bibinfo{author}{\bibfnamefont{J.~K.} \bibnamefont{Nayak}}, \bibnamefont{and}
  \bibinfo{author}{\bibfnamefont{R.}~\bibnamefont{Varma}},
  \bibinfo{journal}{J.Phys.} \textbf{\bibinfo{volume}{G41}},
  \bibinfo{pages}{035106} (\bibinfo{year}{2014}), \eprint{1402.6813}.

\bibitem[{\citenamefont{Aggarwal et~al.}(2011)}]{Aggarwal:2010aa}
\bibinfo{author}{\bibfnamefont{M.}~\bibnamefont{Aggarwal}} \bibnamefont{et~al.}
  (\bibinfo{collaboration}{STAR Collaboration}), \bibinfo{journal}{Phys.Rev.}
  \textbf{\bibinfo{volume}{C83}}, \bibinfo{pages}{064905}
  (\bibinfo{year}{2011}), \eprint{1004.0925}.

\bibitem[{\citenamefont{Shapoval et~al.}(2013)\citenamefont{Shapoval,
  Braun-Munzinger, Karpenko, and Sinyukov}}]{Shapoval:2013jca}
\bibinfo{author}{\bibfnamefont{V.}~\bibnamefont{Shapoval}},
  \bibinfo{author}{\bibfnamefont{P.}~\bibnamefont{Braun-Munzinger}},
  \bibinfo{author}{\bibfnamefont{I.~A.} \bibnamefont{Karpenko}},
  \bibnamefont{and} \bibinfo{author}{\bibfnamefont{Y.~M.}
  \bibnamefont{Sinyukov}}, \bibinfo{journal}{Phys.Lett.}
  \textbf{\bibinfo{volume}{B725}}, \bibinfo{pages}{139} (\bibinfo{year}{2013}),
  \eprint{1304.3815}.

\bibitem[{\citenamefont{Sinyukov and Shapoval}(2013)}]{Sinyukov:2012ut}
\bibinfo{author}{\bibfnamefont{Y.}~\bibnamefont{Sinyukov}} \bibnamefont{and}
  \bibinfo{author}{\bibfnamefont{V.}~\bibnamefont{Shapoval}},
  \bibinfo{journal}{Phys.Rev.} \textbf{\bibinfo{volume}{D87}},
  \bibinfo{pages}{094024} (\bibinfo{year}{2013}), \eprint{1209.1747}.

\bibitem[{\citenamefont{Gubser}(2010)}]{Gubser:2010ze}
\bibinfo{author}{\bibfnamefont{S.~S.} \bibnamefont{Gubser}},
  \bibinfo{journal}{Phys.Rev.} \textbf{\bibinfo{volume}{D82}},
  \bibinfo{pages}{085027} (\bibinfo{year}{2010}), \eprint{1006.0006}.

\bibitem[{\citenamefont{Borsanyi et~al.}(2014)\citenamefont{Borsanyi, Fodor,
  Hoelbling, Katz, Krieg et~al.}}]{Borsanyi:2013bia}
\bibinfo{author}{\bibfnamefont{S.}~\bibnamefont{Borsanyi}},
  \bibinfo{author}{\bibfnamefont{Z.}~\bibnamefont{Fodor}},
  \bibinfo{author}{\bibfnamefont{C.}~\bibnamefont{Hoelbling}},
  \bibinfo{author}{\bibfnamefont{S.~D.} \bibnamefont{Katz}},
  \bibinfo{author}{\bibfnamefont{S.}~\bibnamefont{Krieg}},
  \bibnamefont{et~al.}, \bibinfo{journal}{Phys.Lett.}
  \textbf{\bibinfo{volume}{B730}}, \bibinfo{pages}{99} (\bibinfo{year}{2014}),
  \eprint{1309.5258}.

\bibitem[{\citenamefont{Cooper and Frye}(1974)}]{Cooper:1974mv}
\bibinfo{author}{\bibfnamefont{F.}~\bibnamefont{Cooper}} \bibnamefont{and}
  \bibinfo{author}{\bibfnamefont{G.}~\bibnamefont{Frye}},
  \bibinfo{journal}{Phys.Rev.} \textbf{\bibinfo{volume}{D10}},
  \bibinfo{pages}{186} (\bibinfo{year}{1974}).

\bibitem[{\citenamefont{Hirano et~al.}(2013)\citenamefont{Hirano, Huovinen,
  Murase, and Nara}}]{Hirano:2012kj}
\bibinfo{author}{\bibfnamefont{T.}~\bibnamefont{Hirano}},
  \bibinfo{author}{\bibfnamefont{P.}~\bibnamefont{Huovinen}},
  \bibinfo{author}{\bibfnamefont{K.}~\bibnamefont{Murase}}, \bibnamefont{and}
  \bibinfo{author}{\bibfnamefont{Y.}~\bibnamefont{Nara}},
  \bibinfo{journal}{Prog.Part.Nucl.Phys.} \textbf{\bibinfo{volume}{70}},
  \bibinfo{pages}{108} (\bibinfo{year}{2013}), \eprint{1204.5814}.

\bibitem[{\citenamefont{Pratt}(1986)}]{Pratt:1986cc}
\bibinfo{author}{\bibfnamefont{S.}~\bibnamefont{Pratt}},
  \bibinfo{journal}{Phys.Rev.} \textbf{\bibinfo{volume}{D33}},
  \bibinfo{pages}{1314} (\bibinfo{year}{1986}).

\bibitem[{\citenamefont{Bertsch et~al.}(1988)\citenamefont{Bertsch, Gong, and
  Tohyama}}]{Bertsch:1988db}
\bibinfo{author}{\bibfnamefont{G.}~\bibnamefont{Bertsch}},
  \bibinfo{author}{\bibfnamefont{M.}~\bibnamefont{Gong}}, \bibnamefont{and}
  \bibinfo{author}{\bibfnamefont{M.}~\bibnamefont{Tohyama}},
  \bibinfo{journal}{Phys.Rev.} \textbf{\bibinfo{volume}{C37}},
  \bibinfo{pages}{1896} (\bibinfo{year}{1988}).

\bibitem[{\citenamefont{Kisiel}(2011)}]{Kisiel:2010xy}
\bibinfo{author}{\bibfnamefont{A.}~\bibnamefont{Kisiel}},
  \bibinfo{journal}{Phys.Rev.} \textbf{\bibinfo{volume}{C84}},
  \bibinfo{pages}{044913} (\bibinfo{year}{2011}), \eprint{1012.1517}.

\bibitem[{\citenamefont{Borysova et~al.}(2006)\citenamefont{Borysova, Sinyukov,
  Akkelin, Erazmus, and Karpenko}}]{Borysova:2005ng}
\bibinfo{author}{\bibfnamefont{M.}~\bibnamefont{Borysova}},
  \bibinfo{author}{\bibfnamefont{Y.}~\bibnamefont{Sinyukov}},
  \bibinfo{author}{\bibfnamefont{S.}~\bibnamefont{Akkelin}},
  \bibinfo{author}{\bibfnamefont{B.}~\bibnamefont{Erazmus}}, \bibnamefont{and}
  \bibinfo{author}{\bibfnamefont{I.}~\bibnamefont{Karpenko}},
  \bibinfo{journal}{Phys.Rev.} \textbf{\bibinfo{volume}{C73}},
  \bibinfo{pages}{024903} (\bibinfo{year}{2006}), \eprint{nucl-th/0507057}.

\end{thebibliography}

\end{document}